\documentclass[a4paper,10pt]{article}
\usepackage{amssymb}

\usepackage{graphicx}
\usepackage{amsmath}
\usepackage[english]{babel}
\usepackage{amssymb}
\usepackage{float}
\usepackage{times}
\usepackage{graphicx}
\usepackage{amsmath}
\usepackage{tikz}
\usepackage{color}
\usepackage{graphicx}
\usepackage{rotating}
\usepackage{tikz}
\usepackage{xcolor}
\usepackage[font=small,labelfont={bf}]{caption}
\usepackage{plain}
\usepackage{setspace}
\usepackage{enumerate}
\usepackage{algorithm}
\usepackage{algorithmic}
\usetikzlibrary{arrows}

\def\cg#1{\mbox{${\cal #1}$}}      

\definecolor{dar}{rgb}{0.6,0.2,0.1}

\newtheorem{Theorem}{Theorem}
\newtheorem{Example}{Example}
\newtheorem{Definition}{Definition}
\newtheorem{Corollary}{Corollary}
\newcommand\independent{\protect\mathpalette{\protect\independenT}{\perp}}\def\independenT#1#2{\mathrel{\rlap{$#1#2$}\mkern2mu{#1#2}}}

\begin{document}

\title{Multiple Hidden Markov Models \\for Categorical Time Series}
\vspace{1cm}
\author{\textbf{Roberto Colombi}, \textbf{Sabrina Giordano}\\
[1 cm] Department of Engineering, University of Bergamo, Italy \and
Department of Economics, Statistics and Finance, University of Calabria,
Italy}

\maketitle

\begin{abstract}
\textit{We introduce multiple hidden Markov models (MHMMs) where an observed
multivariate categorical time series depends on an unobservable multivariate Mar- kov chain. MHMMs  provide an elegant framework for specifying various independence relationships between multiple discrete time processes. These independencies are  interpreted as Markov properties of a  mixed graph and a chain graph associated to the latent and observable components of the MHMM, respectively. These Markov properties are also translated into zero restrictions on the parameters of marginal models for the transition probabilities and the distributions of the observable variables given the latent states.}
\end{abstract}
\vspace{0.4cm}


\section{Introduction}
 In several applications involving time series, it is of interest to describe how the evolution of variables over time depends on  latent characteristics or the focus may be on the dynamics of  unobservable characteristics measured by variables observed at consecutive time occasions. These issues are addressed by hidden Markov models and a widespread application of them has occurred in several fields such as speech recognition, signal processing, digital communication, biology, reliability etc., standard references are MacDonald and Zucchini, 1997 and 2009, Capp\'{e} et al., 2005, among others.

Basically, an hidden Markov model assumes that an observable
time series depends on an unobservable Markov chain in such a way
that the joint process
 is also Markovian.

%
%
In this work, we
focus on discrete hidden Markov models  with a multivariate
categorical observable process depending on a multivariate latent chain, so
we observe more variables at each time and assume that their
distribution can be affected by one or more latent variables.  We will refer to these extensions of the traditional hidden Markov process as multiple
hidden Markov models (MHMMs).

MHMMs can be well suited to applications where all the observable time series are affected by one common unobservable factor (general effect) and each observable variable is also governed by its specific latent variable.  MHMMs may also handle time series data where an unobservable aspect influences the marginal dynamics of each observable variable while another latent factor influences the association among them.  In the framework of MHMMs, different sets of observable categorical time series are allowed to depend on different sets of unobservable processes and the observable variables are not required to be independent given the latent states (local independence assumption), but the association between them is also modeled.  Moreover, in MHMMs, the multivariate latent process can satisfy the hypotheses of Granger noncausality and contemporaneous independence described by Colombi and Giordano (2012) for multivariate Markov chains. Therefore, MHMMs widen the opportunities of applications of the classical hidden Markov models. This approach, for example, responds to the shortcomings highlighted in Zucchini and Guttorp, 1991, who proposed a model for describing the sequence of wet and dry days at 5 sites without taking into account the spatial dependence among sites situated in closed locations and without allowing for a multivariate state process with sites in different regions responding to different components of the latent process. Other examples illustrated in the literature can be enriched of more flexible and realistic hypotheses using MHMMs.

In MHMMs, the state space $\mathcal{E}$ is the cartesian product of the state spaces of the single latent variables. These models, without restrictions on the transition probabilities of the latent chain are useless  as they are equivalent to the conventional hidden Markov model (HMM) with a number of states equal to the cardinality of $ \mathcal{E}$.  However, the conventional  HMM has an extremely large number of parameters and cannot be easily  used to discover interesting independence structures in the transition matrix of the latent chain and in the distributions of observable variables given the latent states, while MHMMs enable us to formulate meaningful constraints  offering advantages of parsimony and interpretability.

In the sequel, to make easier the interpretability of the proposed models,  we will associate graphical Markov models to
 MHMMs whose probabilities will be parameterized by marginal models. Thus, our contribution stays also in taking advantages of the use of graphical and marginal models into the framework of hidden Markov processes.

The paper is organized as follows.  MHMMs are presented in Section 2. The transition probabilities of the latent model and the distributions of observable variables given the latent states are required to obey the Markov properties of a  mixed and a chain graph, respectively, in Section 3.
Finally, we will show in Section 4 that the independencies among the observable and latent
variables of the MHMM can be easily verified by testing linear constraints on parameters of  marginal parameterizations (Bergsma and Rudas, 2002) of the transition probabilities and the probabilities of the observations given the latent states.

\section{Multiple hidden Markov models}
Let $\mathbf{E}_{\mathcal{U}}$ be a $r$-variate  process of
categorical variables, $\mathbf{E}_{\mathcal{U}}=$
 $\{E_{\mathcal{U}}(t):t \in
\mathbb{{N}}\}=$  $\{E_{i}(t):t \in \mathbb{N}, i \in \mathcal{U}\}$,
$\mathcal{U}=\{1,...,r\}$, $\mathbb{N}=\{0,1,2,...,\}$ and let
$\mathbf{F}_{\mathcal{V}}$ be a $s$-dimensional process of
categorical variables $\mathbf{F}_{\mathcal{V}}=$
 $\{F_{\mathcal{V}}(t):t \in
\mathbb{{N}}\}=$  $\{F_{j}(t):t \in \mathbb{N}, j \in \mathcal{V}\}$,
$\mathcal{V}=\{1,...,s\}$.  The random variables $E_{i}(t)$, $F_{j}(t)$ take values in
finite sets ${\mathcal E}_i$  and
${\mathcal F}_j$
$i \in \mathcal{U}$,
$ j \in \mathcal{V}$. One realization of the process
$\mathbf{F}_{\mathcal{V}}$ at a given time is denoted by
$\emph{\textbf{f}}=(f_1,f_2,...,f_s) \in \mathcal{F}=\times_{j \in \mathcal{V}}\mathcal{F}_j$, and the realization of  $\mathbf{E}_{\mathcal{U}}$ is
$\emph{\textbf{e}}=(e_1,e_2,...,e_r)$ in $\mathcal{E}=\times_{i \in \mathcal{U}}\mathcal{E}_i$.
For every subset $\mathcal{T} \subset \mathcal{U}$ and $\mathcal{R}
\subset \mathcal{V}$, marginal processes are represented by
$\mathbf{E}_{\mathcal{T}}= \{E_{i}(t): i \in \mathcal{T}, t \in
\mathbb{N}\}$ and $\mathbf{F}_{\mathcal{R}}= \{F_{j}(t): j \in
\mathcal{R}, t \in \mathbb{N}\}.$
Univariate marginal processes will be denoted by
$\mathbf{E}_{i}$, $\mathbf{F}_{j}$, $i \in
\mathcal{U}, j \in \mathcal{V}$.

The following definition states when ($\mathbf{E}_{\mathcal{U}}$, $\mathbf{F}_{\mathcal{V}}$) is an MHMM.
\begin{Definition}
The  joint process
($\mathbf{E}_{\mathcal{U}}$, $\mathbf{F}_{\mathcal{V}}$) is an MHMM if
\begin{itemize}
\item[a)] $\mathbf{E}_{\mathcal{U}}$ is not observable
\item[b)] ($\mathbf{E}_{\mathcal{U}}$, $\mathbf{F}_{\mathcal{V}}$) is a first order multivariate Markov chain
\item[c)] $E_{\mathcal{U}}(t) \independent F_{\mathcal{V}}(t-1)|E_{\mathcal{U}}(t-1)$
\item[d)] $F_{\mathcal{V}}(t)\independent E_{\mathcal{U}}(t-1),F_{\mathcal{V}}(t-1)|E_{\mathcal{U}}(t)$.
\end{itemize}
\end{Definition}
In particular, condition $c)$ implies that $\mathbf{E}_{\mathcal{U}}$ is a first order Markov chain (Colombi and Giordano, 2011).

The marginal process $(\mathbf{E_{\mathcal{U}}},\mathbf{F}_{\mathcal{R}})$, $\mathcal{R} \subset \mathcal{V}$, of  an MHMM ($\mathbf{E}_{\mathcal{U}}$, $\mathbf{F}_{\mathcal{V}}$) is hidden Markov too, whereas
in general  $(\mathbf{E_{\mathcal{T}}},\mathbf{F}_{\mathcal{R}})$, $\mathcal{T} \subset \mathcal{U}$, is not a hidden Markov model.

The following theorem clarifies when the properties of an  MHMM  are preserved after marginalizing  the  latent and observable processes.

  \begin{Theorem} \label{marg} Let $\mathbf{E}_{\mathcal{T}}$ and $\mathbf{F}_{\mathcal{R}}$,  $\mathcal{T} \subset \mathcal{U}$ and $\mathcal{R} \subset \mathcal{V}$, be  marginal processes of the latent and observable components of an MHMM $(\mathbf{E_{\mathcal{U}}},\mathbf{F}_{\mathcal{V}})$. The marginal process $(\mathbf{E_{\mathcal{T}}},\mathbf{F}_{\mathcal{R}})$ is still an MHMM  if and only if the following conditions are satisfied for all $t \in \mathbb{N}\setminus \{0\}$
\begin{equation}\label{M}
E_{\mathcal{T}}(t)
\independent
{E}_{\mathcal{U}\setminus
\mathcal{T}}(t-1)|{E}_{\mathcal{T}}(t-1)
\end{equation}
\begin{equation}
  F_{\mathcal{R}}(t)
\independent
{E}_{\mathcal{U} \setminus
\mathcal{T}}(t)|{E}_{
\mathcal{T}}(t). \label{LI}
\end{equation}
\end{Theorem}

This theorem is a special case of the results presented in Colombi and Giordano (2011) where the statement (\ref{M})
is proved to ensure that
 the marginal process $\mathbf{E}_{\mathcal{T}}$ of the latent component $\mathbf{E_{\mathcal{U}}}$  is still a Markov chain.
Condition (\ref{LI}) states that  observable variables
${F}_{\mathcal{R}}(t)$ at time $t$ depend only on the  latent
variables $E_{\mathcal{T}}(t)$.

\subsection{Special cases of MHMMs: linked and coupled MHMMs}
Here, we discuss some special cases of MHMMs which generalize the
linked  and  coupled hidden Markov models  proposed in the literature in the simple version with two  observable variables and two latent processes (Koski,
2001, Brand et al., 1997).

In our generalized version of linked  hidden Markov models,  every set of a partition of the latent variables is affected only by its own past, while in the coupled  hidden Markov models,  these sets are independent conditionally on the past of every latent variable.
In all the mentioned models, every set of the partition of the latent variables influences one and  only one set of a partition of the manifest variables and  the  sets of the observable variables are independent given the latent states.

A formal definition is presented below.

\begin{Definition} (Linked and coupled MHMMs).\label{Linked}
 The  joint process
($\mathbf{E}_{\mathcal{U}}$, $\mathbf{F}_{\mathcal{V}}$) is a linked MHMM, with $l$ components if and only if there is a partition of the latent variables  $\cg U= \bigcup_{i=1}^{l} \cg T_i$ and
 a partition of the observable variables  $\cg V= \bigcup_{i=1}^{l} \cg R_i$ such that the following conditions are satisfied for
  all the unions of variables $\mathcal{U}_{s}=\bigcup_{i \in  s} \cg T_i$, $\mathcal{V}_{s}=\bigcup_{i \in s} \cg R_i$,  ${s} \subset \{1,2,...,l\}$
\begin{equation} \label{linked}
E_{\mathcal{U}_{s}}(t)
\independent
{E}_{\mathcal{U}\setminus
\mathcal{U}_s}(t-1)|{E}_{\mathcal{U}\mathfrak{}_s}(t-1)
\end{equation}
\begin{equation}\label{obslat}
  F_{\mathcal{V}_{s}}(t)
\independent
{E}_{\mathcal{U} \setminus
\mathcal{U}_s}(t)|{E}_{
\mathcal{U}_{s} }(t)
\end{equation}
\begin{equation}\label{obs}
  F_{\mathcal{R}_{1}}(t)\independent F_{\mathcal{R}_{2}}(t) \independent ...\independent F_{\mathcal{R}_{l}}(t)|{E}_{\mathcal{U}}(t).
\end{equation}
If condition (\ref{linked})
is replaced by
\begin{equation}\label{coupled}
  E_{\mathcal{T}_{1}}(t)\independent E_{\mathcal{T}_{2}}(t) \independent ...\independent E_{\mathcal{T}_{l}}(t)|{E}_{\mathcal{U}}(t-1).
\end{equation}
the process ($\mathbf{E}_{\mathcal{U}}$, $\mathbf{F}_{\mathcal{V}}$) satisfying (\ref{obslat}), (\ref{obs}) and (\ref{coupled}) is called coupled MHMM.
\end{Definition}

 It is worthwhile to note that according to Theorem \ref{marg}, the marginal processes ($\mathbf{E}_{\mathcal{U}_s}$, $\mathbf{F}_{\mathcal{V}_{s}
})$,   $s \subset \{1,2,...,l\}$, of a linked MHMM
are still MHMMs, but this is not true for a coupled MHMM.

 We stress that it is important to involve the unions of variables $\mathcal{U}_{s}=\bigcup_{i \in  s} \cg T_i$, $\mathcal{V}_{s}=\bigcup_{i \in s} \cg R_i$ in the conditions (\ref{linked}) and (\ref{obslat}). In fact, if they were replaced  by the weaker version $E_{\mathcal{T}_{i}}(t)
\independent
{E}_{\mathcal{U}\setminus
\mathcal{T}_i}(t-1)|{E}_{\mathcal{T}\mathfrak{}_i}(t-1)$ and $ F_{\mathcal{R}_{i}}(t)
\independent
{E}_{\mathcal{U} \setminus
\mathcal{T}_i}(t)|{E}_{
\mathcal{T}_{i} }(t)$ with  $i=1,2,...,l$,
it could happen, for example, that $\mathbf{E}_{\mathcal{T}_i}$ and  $\mathbf{E}_{\mathcal{T}_j}$ are marginally affected only by their own past, but $\mathbf{E}_{\mathcal{T}_i \cup \mathcal{T}_j}$ is influenced by the past of all the latent variables.
 Similarly, it could happen that $\mathbf{F}_{\mathcal{R}_i}$ and $\mathbf{F}_{\mathcal{R}_j}$  respond marginally only to $\mathbf{E}_{\mathcal{T}_i}$ and  $\mathbf{E}_{\mathcal{T}_j}$, respectively, but their joint distribution depends on all the latent variables. Moreover in linked MHMM, it could be that the marginal processes $(\mathbf{E}_{\mathcal{T}_i},\mathbf{F}_{\mathcal{R}_i})$ and $(\mathbf{E}_{\mathcal{T}_j}, \mathbf{F}_{\mathcal{R}_j})$ are hidden Markov, but
 $(\mathbf{E}_{\mathcal{T}_i \cup \mathcal{T}_j},\mathbf{F}_{\mathcal{R}_i \cup \mathcal{R}_j})$ is no longer an MHMM.

\section{Graphical models for MHMMs}

In this section, we address the use of
graphical Markov models for MHMMs. Such models associate missing edges of a graph with some conditional independence restrictions imposed on
the probabilities of the observable variables given the latent states and the transition probabilities of the latent process.

The terminology and the notation used throughout the paper follow Colombi and Giordano, 2012. We just briefly remind that $G=(V,E)$ is a graph with a finite set of nodes $V$ and a set of edges $E$; moreover, for every non-empty subset of nodes $\mathcal{S}$,
$pa_{G}(\mathcal{S})$, $ch_{G}(\mathcal{S})$, $nb_{G}(\mathcal{S})$, and $sp_{G}(\mathcal{S})$ are the
collection of parents, children, neighbours and spouses of nodes in the set $\mathcal{S}$. In particular, here the nodes can correspond to variables or to processes.
Note that  every node is neighbour and spouse of itself.

Finally, $\mathfrak{B}(G)$ indicates the family of the bi-connected sets in the graph.

\subsection{MHMM Markov with respect to  a mixed and a chain graph}
We consider two types of graphs: a mixed graph for the latent component of the MHMM and a chain graph for the observable component.
In particular, the transition probabilities of the multivariate latent Markov chain $\mathbf{E}_{\mathcal{U}}$ in the MHMM are required to obey a set of Markov properties encoded by a mixed graph $G$  whose basic features are discussed  in Colombi and Giordano, 2012, for Markov chains, while the Markov properties satisfied by the distribution of the observable variables conditioned on the latent states are read off a chain graph $G^*$.

In a mixed graph $G$, a  node $i$ corresponds to the marginal process $\mathbf{E}_{i}$, for every $i \in \mathcal{U}$, and
independence restrictions are associated with
missing directed and bi-directed edges, respectively. In particular,
missing bi-directed edges lead to independencies of marginal processes
at the same point in time; missing directed edges, instead, refer to
independencies which involve marginal processes at two consecutive
instants.

A chain graph $G^*$ with two
chain components $\tau_0$
  and $\tau_1$ serves the need to encode the independence relations among observable and latent variables of the MHMM at a given time point.
 The nodes  of the chain graph $G^*$,  belonging to $\tau_0$ correspond to the random variables $E_i(t^*)$, $i \in \mathcal{U}$, and  the nodes belonging to $\tau_1$,
correspond to the random variables  $F_j(t^*)$,
$j \in \mathcal{V}$, for any arbitrary $t^* \in \mathbb{N}$.

All the edges in the subgraph induced by a chain component are
bi-directed and the graph induced by the chain component
$\tau_0$ is bi-complete. Furthermore, the directed edges in graph
$G^*$ point in the same direction from $\tau_0$ towards $\tau_1$.

The definition below specifies the properties for an MHMM being Markov with respect to a mixed and a chain graph.

\begin{Definition} (MHMM Markov wrt a mixed and a chain graph).\label{mixchain} The latent process $\mathbf{E}_{\mathcal{U}}$ of an MHMM ($\mathbf{E}_{\mathcal{U}}$,$\mathbf{F}_{\mathcal{V}}$) is Markov wrt a mixed graph $G$ if and only if its transition
probabilities satisfy the following
 conditional independencies  for all $t \in \mathbb{N}\setminus \{0\}$ associated to missing directed and bi-directed edges of $G$, respectively:
\begin{eqnarray}
E_{\mathcal{T}}(t) \independent  E_{\mathcal{U} \backslash
pa_{G}(\mathcal{T})}(t-1)|E_{pa_{G}(\mathcal{T})}(t-1) \hspace{1 cm} \forall
\mathcal{T} \subset \mathcal{U} \label{c3}\\[0.2 cm]
E_{\mathcal{T}}(t) \independent  E_{\mathcal{U} \backslash
sp_{G}(\mathcal{T})}(t)|E_{\mathcal{U}}(t-1) \hspace{1 cm} \forall
\mathcal{T} \subset \mathcal{U}. \label{c2}
\end{eqnarray}
The observable process $\mathbf{F}_{\mathcal{V}}$ is Markov wrt a chain graph $G^*$ if and only if the distribution of the observable variables given the latent states satisfies the following
 conditional independencies  for all $t \in \mathbb{N}\setminus \{0\}$ associated to missing bi-directed  and directed edges of $G^*$, respectively:
\begin{eqnarray}F_{\mathcal{R}}(t) \independent  F_{\mathcal{V} \backslash
sp_{G^*}(\mathcal{R})}(t)|E_{\mathcal{U}}(t) \hspace{1 cm} \forall
\mathcal{R} \subset \mathcal{V} \label{c2b}\\[0.2 cm]
F_{\mathcal{R}}(t) \independent  E_{\mathcal{U} \backslash
pa_{G^*}(\mathcal{R})}(t)|E_{pa_{G^*}(\mathcal{R})}(t) \hspace{1 cm} \forall
\mathcal{R} \subset \mathcal{V}. \label{c3b}
\end{eqnarray}
Therefore, an MHMM ($\mathbf{E}_{\mathcal{U}}$,$\mathbf{F}_{\mathcal{V}}$) is said to be Markov wrt a  mixed and a chain graph when the latent component is Markov
wrt a mixed graph, and the observation component given the latent states is Markov wrt a chain graph.
\end{Definition}
\vspace{0.3 cm}

 In the context of first order multivariate Markov chains, condition (\ref{c3}) corresponds to a notion of Granger noncausality and, when $pa_G(\mathcal{T})=\mathcal{T}$, it ensures that the marginal process $\mathbf{E}_{\mathcal{T}}$ is a Markov chain (Colombi and Giordano, 2012, Florens et al., 1993). Henceforth, we will refer to (\ref{c3}) with the term Granger
noncausality (G-noncausality) condition saying that the latent process ${\mathbf
E}_{\mathcal{T}}$ is not G-caused by ${\mathbf E}_{\mathcal{U}
\backslash pa_G(\mathcal{T})}$ with respect to $\mathbf{E}_{\mathcal{U}}$.
Condition (\ref{c2})  states that the
transition probabilities satisfy the bi-directed Markov
property (Richardson, 2003) with respect to the graph obtained by
removing the directed edges from the mixed graph. Here, we will refer to
(\ref{c2}) with the term contemporaneous independence
condition  and say that the latent processes ${\mathbf E}_{\mathcal{T}}$
and  ${\mathbf E}_{\mathcal{U} \backslash sp_G(\mathcal{T})}$ are
contemporaneously independent.

On the other hand, conditions (\ref{c2b}) and (\ref{c3b}) encoded by the chain graph $G^*$ refer to the observable component of the MHMM  and are equivalent to the type IV Markov properties $C2b$ and $C3b$ discussed by Drton, 2009. Note that conditions (\ref{c2b}) are  bi-directed Markov properties which describe a local independence assumption,  the independencies (\ref{c3b}) of a set of manifest variables from a set of latent variables, at time $t$, are conditioned on the remaining latent variables, but not to the remaining observable variables.

 Missing edges in $G^*$ can be alternatively interpreted as the  type I Markov properties $C2a$ and $C3a$ proposed by Drton, 2009.
 In such a case,  bi-directed edges in $G^*$ are usually replaced by undirected edges and the Markov properties expressed by the following statements for all $t \in \mathbb{N}\setminus \{0\}$
\begin{eqnarray*}
F_{\mathcal{R}}(t) \independent  F_{\mathcal{V} \backslash
nb_{G^*}(\mathcal{R})}(t)|E_{\mathcal{U}}(t),F_{nb_{G^*}(\mathcal{R}) \backslash \mathcal{R}}(t) \hspace{1 cm} \forall
\mathcal{R} \subset \mathcal{V} \label{c2bis}\\[0.2 cm] F_{\mathcal{R}}(t) \independent  E_{\mathcal{U} \backslash
pa_{G^*}(\mathcal{R})}(t)|E_{pa_{G^*}(\mathcal{R})}(t),F_{nb_{G^*}(\mathcal{R})}(t) \hspace{1 cm} \forall
\mathcal{R} \subset \mathcal{V}. \label{c3bis}
\end{eqnarray*}
Although this type I chain graph model is simpler since it corresponds to zero constraints on standard log-linear parameters of the observation probabilities, in our opinion the independence conditions it encodes are less meaningful than type IV Markov properties. In fact, when the independence of set of observable variables from some latent variables is of interest it seems inappropriate to condition also with respect to the remaining observable variables.

\begin{figure}[h!]
\centering \resizebox{0.8 \textwidth}{!}
{\includegraphics{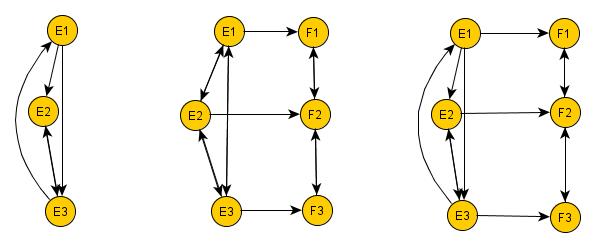}} \caption{Mixed (left), chain graph (middle) and the mixed-chain graph (rigth)  associated to an MHMM}
\label{figrs0}
\end{figure}

As a matter of convenience, the mentioned mixed graph and the two component chain graph can be superimposed as shown in Figure \ref{figrs0} to form a unique graph  that we will refer to as mixed-chain graph for simplicity.

In all the examples throughout the paper, we implicitly assume that each node of the mixed-chain graph, corresponding to a latent variable, is parent of itself, even if the  edge $\mathbf{\circlearrowleft}$ is not reported.

\begin{figure}[h!]
\centering \resizebox{0.75 \textwidth}{!}
{\includegraphics{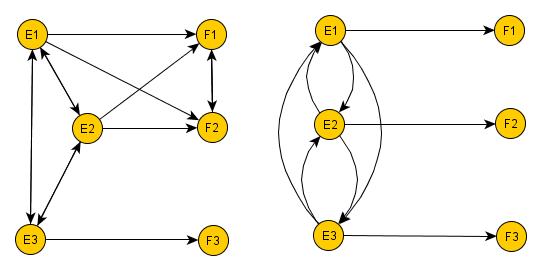}} \caption{Mixed-chain graphs associated to MHMMs}
\label{figrs1}
\end{figure}

\begin{Example} The mixed-chain graph on the right of Figure \ref{figrs1} encodes that the latent variables of the three dimensional MHMM are Granger caused reciprocally but are contemporaneously independent, i.e. $E_1(t) \independent  E_2(t) \independent E_3(t)|E_{\{1,2,3\}}(t-1)$; moreover, every observable variable depends only on its own latent variable,  i.e. $F_1(t) \independent  E_{\{2,3\}}(t)|E_1(t)$, $F_2(t) \independent  E_{\{1,3\}}(t)|E_2(t)$, $F_3(t) \independent  E_{\{1,2\}}(t)|E_3(t)$, $F_{\{1,2\}}(t)$ $ \independent  E_3(t)|E_{\{1,2\}}(t)$, $F_{\{2,3\}}(t) \independent  E_1(t)|E_{\{2,3\}}(t)$, $F_{\{1,3\}}(t) \independent  E_2(t)|E_{\{1,3\}}(t)$;  and, at every time point, the observable variables given the latent states are independent, i.e. $F_1(t) \independent  F_2(t) \independent F_3(t)|E_{\{1,2,3\}}(t)$.

The mixed-chain graph on the left of Figure \ref{figrs1} is associated to a 3-variate MHMM where the latent variables are not contemporaneously independent, and each of them depends only on its own past, i.e. the encoded Granger noncausality conditions (\ref{c3}) are $E_1(t) \independent  E_{\{2,3\}}(t-1) |E_1(t-1)$, $E_2(t) \independent  E_{\{1,3\}}(t-1) |E_2(t-1)$, $E_3(t) \independent  E_{\{1,2\}}(t-1)|E_3(t-1)$, $E_{\{1,2\}}(t) \independent  E_3(t-1) |E_{\{1,2\}}(t-1)$, $E_{\{2,3\}}(t) \independent  E_1(t-1) |E_{\{2,3\}}(t-1)$, $E_{\{1,3\}}(t) \independent  E_2(t-1) |E_{\{1,3\}}(t-1)$;  while the independence conditions (\ref{c3b}) of manifest  from latent variables are $F_{\{1,2\}}(t) \independent  E_3(t) |E_{\{1,2\}}(t)$ and $F_3(t) \independent  E_{\{1,2\}}(t) |E_3(t)$; the local independence (\ref{c2b}) is $F_1(t) \independent  F_{\{2,3\}}(t) |E_{\{1,2,3\}}(t)$.
\end{Example}

\begin{figure}[h!]
\centering \resizebox{0.7 \textwidth}{!}
{\includegraphics{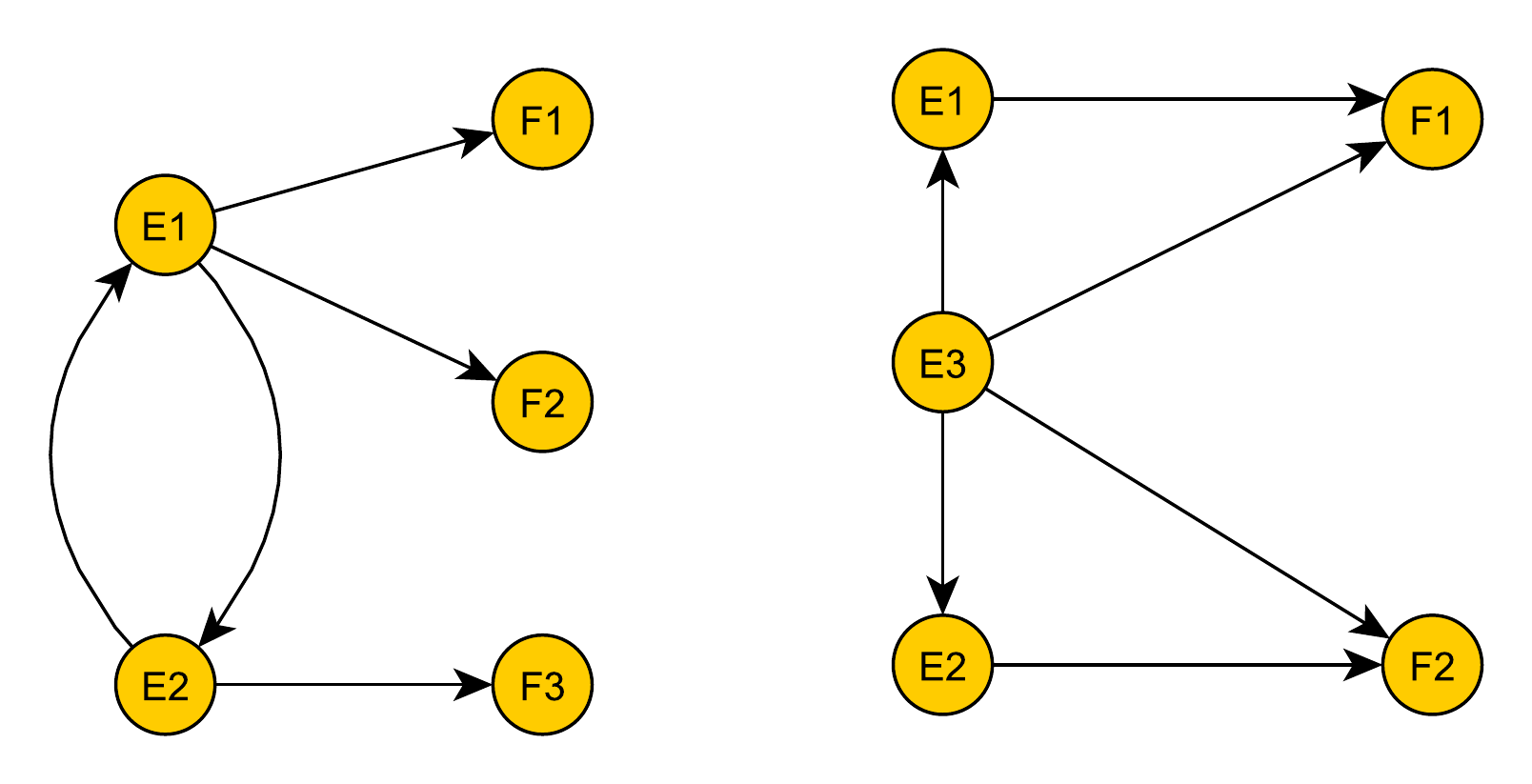}} \caption{Mixed-chain graphs associated to MHMMs for marketing and financial data illustrated in Example 2}
\label{figrsex}
\end{figure}

\begin{Example} (MHMMs for marketing and financial data). The mixed-chain graph on the left in Figure \ref{figrsex} is associated to an MHMM encoding the dynamic relations of three observable time series which depend on two latent factors. The observable time series
can represent, for example, the sales levels of three products that can be interpreted as indicators of the motivational latent states of the customers behaviors. The distributions of the motivational states may be of interest more than those of the product sales.
The graph underlies that  the first two products belong to the same category so their sales series ($F_1, F_2$) depend on the latent variable $E_1$ $(i.e.\ F_{\{1,2\}}(t) \independent  E_2(t) |E_1(t))$, the sales of the third product respond to a different latent variable $E_2$ $(i.e.\ F_3(t) \independent  E_1(t) |E_2(t))$; the observable sales levels are independent given the latent states  $(i.e.\ F_1(t) \independent  F_2(t) \independent F_3(t)|E_{\{1,2\}}(t))$;  the two motivational latent processes are G-caused reciprocally, so that the past motivational states influence the actual state of each latent variable, but they are contemporaneously independent $(i.e.\ E_1(t) \independent  E_2(t) |E_{\{1,2\}}(t-1))$.

The mixed-chain graph on the right in Figure \ref{figrsex} can describe a model for financial series where specific and generic latent effects exist.  The series $F_1, F_2$ can indicate the trading patterns of two financial traded shares in two different financial sectors. The presence or the absence of trading may depend on a specific latent aspect typical of each financial sector where the trading takes place (e.g. the turbulence of the sector), so each observable variable depends on one specific latent variable: $F_1$ on $E_1$, $F_2$ on $E_2$, but all the trading patterns may be influenced by one common unobservable variable ($E_3$) such as the Market turbulence with states that correspond  to calm or turbulent phases of the Market $(i.e.\ F_1(t) \independent  E_2(t) |E_{\{1,3\}}(t), F_2(t) \independent  E_1(t) |E_{\{2,3\}}(t))$. There is local independence $F_1(t) \independent  F_2(t)|E_{\{1,2,3\}}(t)$. Moreover, the past Market turbulence can affect the current turbulence of each financial sector (G-causality) whereas a specific latent variable  of a sector does not Granger cause the other specific latent variable and the turbolence of the Market, $(i.e.\ E_3(t) \independent  E_{\{1,2\}}(t-1) |E_3(t-1), E_{\{1,3\}}(t) \independent  E_2(t-1) |E_{\{1,3\}}(t-1), E_{\{2,3\}}(t) \independent  E_1(t-1) |E_{\{2,3\}}(t-1))$. Finally there is no contemporaneous relation among the three turbulences $(i.e.\ E_1(t) \independent  E_2(t) \independent  E_3(t)|E_{\{1,2,3\}}(t-1))$.
\end{Example}

Two final remarks can be advisable.
First, one immediate consequence of Theorem \ref{marg} and Definition \ref{mixchain} is the following result
\begin{Corollary}\label{cor1}
If the  MHMM $(\mathbf{E_{\mathcal{U}}},\mathbf{F}_{\mathcal{V}})$  is Markov wrt a mixed graph $G$ and a chain graph $G^*$, such that  $pa_{G}(\mathcal{T})=\mathcal{T}$ and $pa_{G^*}(\mathcal{R})= \mathcal{T}$,  then the marginal process $(\mathbf{E_{\mathcal{T}}},\mathbf{F}_{\mathcal{R}})$, $\mathcal{T} \subset \mathcal{U}$, $\mathcal{R} \subset \mathcal{V}$, is an MHMM.
\end{Corollary}
\emph{Proof.} Conditions (\ref{c3}) and (\ref{c3b}) for $(\mathbf{E_{\mathcal{U}}},\mathbf{F}_{\mathcal{V}})$ being Markov with respect to the graphs where $pa_{G}(\mathcal{T})=\mathcal{T}$ and $pa_{G^*}(\mathcal{R})= \mathcal{T}$ coincide with the conditions (\ref{M}) and (\ref{LI}) ensuring that $(\mathbf{E_{\mathcal{T}}},\mathbf{F}_{\mathcal{R}})$,$\mathcal{T} \subset \mathcal{U}$, $\mathcal{R} \subset \mathcal{V}$, is still a hidden process.

\vspace{0.5 cm}

Moreover, the restrictions which the observation and latent models obey in linked and coupled MHMMs (Definition 2) can be described by the Markov properties of mixed and chain graphs as specified in the corollary below.

\begin{Corollary}\label{cor2}
An MHMM $(\mathbf{E_{\mathcal{U}}},\mathbf{F}_{\mathcal{V}})$ which is Markov wrt a mixed graph $G$ and a chain graph $G^*$ is a linked MHMM with $l$ components if, for a
   partition of the latent variables $\cg U= \bigcup_{i=1}^{l} \cg T_i$ and a partition of
 of the observable variables  $\cg V= \bigcup_{i=1}^{l} \cg R_i$, $i=1,...,l$, it holds that: $pa_{G}(\mathcal{T}_{i})=\mathcal{T}_{i}$,  $pa_{G^*}(\mathcal{R}_{i})= \mathcal{T}_{i}$ and $sp_{G^*}(\mathcal{R}_{i})= \mathcal{R}_{i}.$
 If the condition $pa_{G}(\mathcal{T}_{i})=\mathcal{T}_{i}$ is replaced by $sp_{G}(\mathcal{T}_{i})=\mathcal{T}_{i}$, the process $(\mathbf{E_{\mathcal{U}}},\mathbf{F}_{\mathcal{V}})$ Markov wrt such graphs is a coupled MHMM.
\end{Corollary}
The proof is omitted as similar to that of Corollary 1.

For instance, in Figure \ref{figrs1}, the mixed-chain graph on the left corresponds to a linked MHMM with two components and the mixed-chain graph on the right to a coupled MHMM with three components.

\subsection{Equivalent mixed-chain graphs}
Standard hidden Markov models, and more in general mixture models, suffer from the non-identifiability problem due to the invariance with respect to  relabeling of hidden states. For this reason, MHMMs are identifiable only up to switching the latent categories and the labels of the latent variables.
Due to the invariance with respect to switching labels of the latent variables, there are mixed-chain graphs equivalent in the sense that they correspond to the same MHMM. We now clarify this point thoroughly.

Two mixed-chain graphs, say $G_1$ and $G_2$, with the same sets of nodes, are equivalent if there is a bijection $\nu$ which maps the set of nodes corresponding to the latent variables onto itself assuring that: \emph{i}) the nodes $E_i$ of $G_1 $and $\nu(E_i)$ of $G_2$ are associated to  latent variables with the same number of states; \emph{ii}) if $E_i$, $E_j$ are connected by an edge in $G_1$ then $\nu(E_i)$ and $\nu(E_j)$ are connected by the same type of edge in $G_2$; \emph{iii}) if $E_i$, $E_j$ are not connected by any edges in $G_1$ then $\nu(E_i)$ and $\nu(E_j)$ are not connected in $G_2$; \emph{iv}) if $E_i$ is a parent of $F_j$ in $G_1$ then $\nu(E_i)$ is a parent of $F_j$ in $G_2$.

It is evident that two mixed-chain graphs, equivalent according to the previous conditions, encode conditional independencies that differ only for the labels assigned to the latent variables.

An example  illustrates the mentioned concepts.

\begin{figure}[h!]
\centering \resizebox{0.8 \textwidth}{!}
{\includegraphics{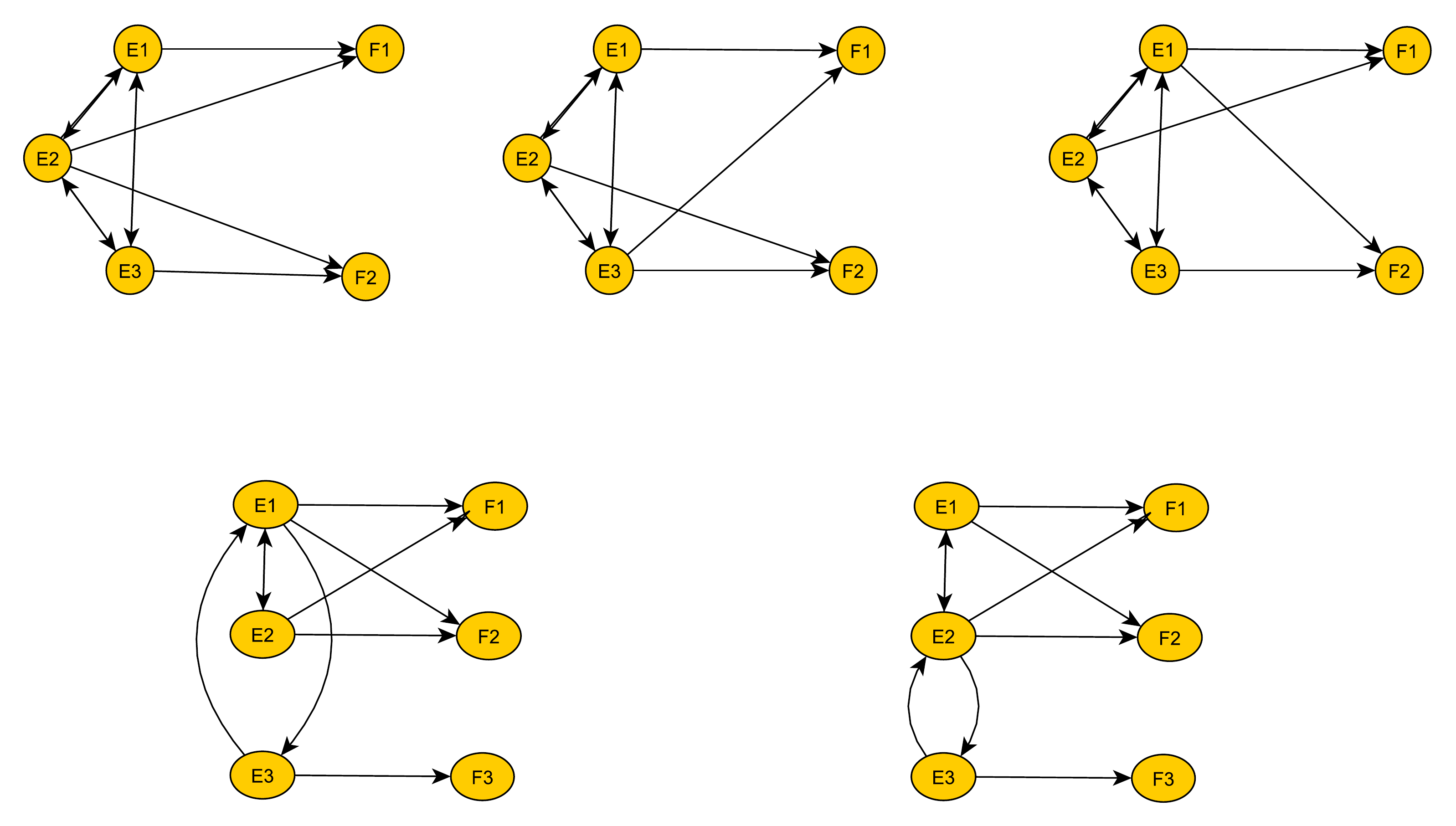}}
\caption{Equivalent mixed-chain graphs
}
\label{figequiv}
\end{figure}

\begin{Example}
The first line of Figure \ref{figequiv} shows three equivalent mixed-chain graphs corresponding to an MHMM where a latent variable affects both the observable variables (generic latent effect) and each manifest variable is governed by its specific latent variable  (specific latent effect). In the second line of Figure \ref{figequiv}, the two equivalent mixed-chain graphs identify the MHMM where two latent variables affect only the first two observable variables and another latent variable is specific for the third observable variable. Moreover, the latent variable specific for $F_3$ is contemporaneously independent of the other two latent variables, and there is no Granger-causality between this variable and one of the remaining unobservable variables.
\end{Example}

\section{A parameterization for MHMMs} \label{McCullagh}

Till now we have described the hypotheses underlying MHMMs, now we need to illustrate how to test such hypotheses. In particular, all the conditional independencies so far illustrated are related to the natural parameters of the MHMMs: the
probabilities of the observable variables given the latent states
and the transition probabilities  of the latent process.

The time-homogeneous joint transition probabilities are denoted by
$\phi(\emph{\textbf{\text{e}}} |\emph{\textbf{\text{e}}}')$ for
every pair of states $\emph{\textbf{\text{e}}}'  \in \mathcal{E}$,
$\emph{\textbf{\text{e}}} \in \mathcal{E}$. Moreover,
$\phi_\mathcal{T}(\emph{\textbf{\text{e}}}_\mathcal{T}
|\emph{\textbf{\text{e}}}')$ is the marginal transition probability
from state $\emph{\textbf{\text{e}}}'  \in \mathcal{E}$ to state
$\emph{\textbf{\text{e}}}_\mathcal{T}$ with components $e_i: i\in
\mathcal{T} \subset \mathcal{U}$.
We assume that the
initial distribution of $\mathbf{E}_{\mathcal{U}}$ is the invariant
one and is unique. Finally,
$\varphi(\emph{\textbf{f}}|\emph{\textbf{e}})$  indicates the state-dependent distribution that is the  conditional probabilities of the observable variables given
the latent state $\textbf{\emph{e}}$ and
$\varphi_{\mathcal{R}}(\emph{\textbf{f}}_\mathcal{R}|\emph{\textbf{e}})$
represents the marginal  probability of the observable variables in the
set $\mathcal{R}$ given the latent state $\textbf{\emph{e}}$.

In this section, we propose to parameterize both the state-dependent distributions
and the transition probabilities  of the latent process through marginal models and  verify hypotheses which make these models more parsimonious by constraining marginal parameters. In the marginal models, the  parameters are
called {\it marginal interactions}. In particular, marginal
interactions are log-linear parameters defined in different
marginal distributions in Bergsma and Rudas (2002) models, while Bartolucci
{\it et al.}, 2007, use more general marginal interactions  which
are  meaningful when the variables have an ordinal nature.
Here, we adopt a Gloneck-McCullagh multivariate logistic model (Gloneck and McCullagh, 1995) whose
interactions, involving the variables in the set $\mathcal{P}$, are log-linear parameters defined on the marginal distributions $\phi_\mathcal{P}(\textbf{\emph{e}}_\mathcal{P}
|\textbf{\emph{e}}')$, $\varphi_{\mathcal{P}}(\textbf{\emph{f}}_\mathcal{P}|\textbf{\emph{e}})$.

For every observable or latent categorical variable, the first
category is called baseline. Any observation
$\textbf{\emph{f}}=(f_1,f_2,..., f_s) $ which includes categories
at the baseline level for variables $j
\notin \mathcal{J}$, $\mathcal{J }\subset \mathcal{V},$  is denoted by
$(\emph{\textbf{\text{f}}}_{\mathcal{J}},\emph{\textbf{\text{f*}}}_{\mathcal{{V}}
\setminus \mathcal{J}})$. A similar notation holds for the latent state $
\textbf{\emph{e}}=(e_1,e_2,...,e_r)$.
For every non-empty subset $\mathcal{P}$ of the observable
variables $\mathcal{V}$ and for every
$\emph{\textbf{\text{f}}}_{\mathcal{P}} \in \times_{j \in
\mathcal{P}}\mathcal{F}_j$, the
 baseline interactions
$\eta^{\mathcal{P}}(\emph{\textbf{\text{f}}}_{
\mathcal{P}}|\emph{\textbf{\text{e}}})$, $\mathcal{P} \in \mathcal{V}$, of the Gloneck-McCullagh marginal model for the
observable variables are
 contrasts of logarithms of marginal
state-dependent probabilities
\begin{equation*}
 \eta^{\mathcal{P}}(\emph{\textbf{\text{f}}}_{\mathcal{P}}|\emph{\textbf{\text{e}}})= \sum_{\mathcal{K}\subseteq \mathcal{P}} (-1)^{|
\mathcal{P}\backslash \mathcal{K}|} \log \varphi_{
\mathcal{P}}(\emph{\textbf{\text{f}}}_{
\mathcal{K}},\emph{\textbf{\text{f*}}}_{\mathcal{P}
\setminus \mathcal{K}}|\emph{\textbf{\text{e}}}).\label{defrob}
\end{equation*}
In order to model the dependence of the distribution of the
observable variables on the states
$\emph{\textbf{\text{e}}}$, we adopt the usual factorial expansion
\begin{equation}\label{teta}
\eta^{\mathcal{P}}(\emph{\textbf{\text{f}}}_{
\mathcal{P}}|\emph{\textbf{\text{e}}}) =\sum_{\mathcal{Q}\subseteq
\mathcal{U}}
\theta^{\mathcal{P},\mathcal{Q}}(\emph{\textbf{\text{f}}}_{
\mathcal{P}}|\emph{\textbf{\text{e}}}_{\mathcal{Q}}).
\end{equation}


Analogously, in the
marginal model for the latent component of MHMMs, we define, for
every $\mathcal{P} \subseteq \mathcal{U}$, the
marginal parameters
\begin{equation*}
 \lambda^{\mathcal{P}}(\emph{\textbf{\text{e}}}_\mathcal{P}|\emph{\textbf{\text{e}}}')= \sum_{\mathcal{K}\subseteq \mathcal{P}} (-1)^{|
\mathcal{P}\backslash \mathcal{K}|} \log \phi_{\mathcal{P}}(\emph{\textbf{\text{e}}}_\mathcal{P} |\emph{\textbf{\text{e}}}').\label{defrob}
\end{equation*}
and the factorial expansion
\begin{eqnarray}
\lambda^{\mathcal{P}}(\emph{\textbf{\text{e}}}_{
\mathcal{P}}|\emph{\textbf{\text{e}}}') =\sum_{\mathcal{Q}\subseteq
\mathcal{U}}
\delta^{\mathcal{P},\mathcal{Q}}(\emph{\textbf{\text{e}}}_{
\mathcal{P}}|\emph{\textbf{\text{e}}}'_{\mathcal{Q}}).\label{logconstr0b}
\end{eqnarray}

\subsection{Parametric constraints for conditional independencies}

The properties  of graphical models for MHMMs (Definition \ref{mixchain}) correspond
 to zero restrictions on the parameters
$\theta^{\mathcal{P},\mathcal{Q}}(\emph{\textbf{{f}}}_{
\mathcal{P}}|\emph{\textbf{{e}}}_{\mathcal{Q}})$ and $\delta^{\mathcal{P},\mathcal{Q}}(\emph{\textbf{{e}}}_{
\mathcal{P}}|\emph{\textbf{{e}}}'_{\mathcal{Q}})$, introduced in (\ref{teta}) and (\ref{logconstr0b}), as illustrated in the next theorem.

\begin{Theorem} \label{param} For a latent model with strictly positive time-homogeneous transition probabilities,
the Granger noncausality condition  (\ref{c3})
is equivalent to
 $\delta^{\mathcal{P},\mathcal{Q}}({\textbf{{e}}}_{
\mathcal{P}}|{\textbf{{e}}}'_{\mathcal{Q}})=\mathbf{0}$
 for all $\mathcal{P} \subseteq \mathcal{T}, \mathcal{Q} \not \subseteq pa_G(\mathcal{P})$, while the conditional contemporaneous independence (\ref{c2})
 is equivalent to $\delta^{\mathcal{P},\mathcal{Q}}({\textbf{{e}}}_{
\mathcal{P}}|{\textbf{{e}}}'_{\mathcal{Q}})=\mathbf{0}$
 for all $\mathcal{P} \not \in \mathfrak{B}(G)$, $\textbf{{e}}_{\mathcal{P}} \in\times_{i \in \mathcal{P}}\mathcal{E}_i$,
$\textbf{{e}}'_{\mathcal{Q}} \in\times_{i \in \mathcal{Q}}\mathcal{
E}_i$.

Moreover, if the state-dependent probabilities are strictly
positive,
the independence (\ref{c2b})
 is equivalent to
$\theta^{\mathcal{P},\mathcal{Q}}({\textbf{{f}}}_{
\mathcal{P}}|{\textbf{{e}}}_{\mathcal{Q}})=\mathbf{0}$ for all
$\mathcal{P} \not \in \mathfrak{B}(G^*)$, while condition (\ref{c3b}) corresponds to $\theta^{\mathcal{P},\mathcal{Q}}({\textbf{{f}}}_{
\mathcal{P}}|{\textbf{{e}}}_{\mathcal{Q}})=\mathbf{0}$ for all
$\mathcal{P} \subseteq \mathcal{R}$, $\mathcal{Q} \not \subseteq pa_{G^*}(\mathcal{P})$, $\textbf{{f}}_\mathcal{P} \in\times_{j \in
\mathcal{P}}\mathcal{F}_j$, $\textbf{{e}}_{\mathcal{Q}}\in\times_{i \in
\mathcal{Q}}\mathcal{E}_i$.
\end{Theorem}

\emph{Proof.} The equivalence between the zero restrictions on $\delta$ parameters and conditions (\ref{c3}) and (\ref{c2}) follows from Colombi and Giordano, 2012, the nullity of $\theta$ parameters under the conditions (\ref{c2b}) and (\ref{c3b}) from Marchetti and Lupparelli, 2011.

\vspace{0.5 cm}

The theorem allows a simple implementation of standard methods to fit and test MHMMs under the restrictions (\ref{c3}, \ref{c2}, \ref{c2b}, \ref{c3b}). The theorem enhances the use of the marginal parametrization since all the conditions (\ref{c3}, \ref{c2}, \ref{c2b}, \ref{c3b}) are equivalent to linear constraints on the marginal parameters, the same restrictions under the log-linear parametrization would correspond to non linear constraints on the parameters.

Moreover, since Corollary \ref{cor1} proves that conditions (\ref{M}) and (\ref{LI})
are particular cases of the
independencies (\ref{c3}) and (\ref{c3b}), the  constraints on marginal parameters described  in Theorem \ref{param} are useful
also for testing if the properties of an  MHMM  are preserved after marginalizing  the  latent and observable processes.

An example clarifies which marginal parameters are restricted to zero in order to satisfy the conditional independencies  that correspond to the Markov properties of the  mixed-chain graph.

\begin{Example} The left mixed-chain graph of Figure \ref{figese3} encodes the following Markov properties: $E_{1}(t) \independent  E_{2}(t-1)|E_{1}(t-1)$ so that $E_1$ is no Granger caused by $E_2$  and $F_{i}(t) \independent  E_{j}(t)|E_{i}(t)$ for $i,j=1,2,$ $i \neq j$ which reveals a specific effect of the first (second) latent variable on the first (second) observable variable.
According to Theorem \ref{marg}, these conditions ensure that the marginal process $(\mathbf{E}_{{1}}$, $\mathbf{F}_{{1}})$ of the MHMM is still hidden Markov, but $(\mathbf{E}_{{2}}$, $\mathbf{F}_{{2}})$ is not.
Moreover, the local independence condition $F_{1}(t) \independent  F_{2}(t)|E_{\{1,2\}}(t)$ can be also read off the same graph. The mentioned conditional independencies  are equivalent to the nullity of the following parameters: $\theta^{1,2}({{{f}}}_{
1}|{{{e_2}}})$, $\theta^{1,12}({{{f}}}_{
1}|{{{e_1,e_2}}})$, $\theta^{2,1}({{{f}}}_{
2}|{{{e_1}}})$, $\theta^{2,12}({{{f}}}_{
2}|{{{e_1,e_2}}})$, $\theta^{12}({{{f}}}_{
1},{{{f}}}_{
2})$ $\theta^{12,1}({{{f}}}_{
1},{{{f}}}_{
2}|{{{e_1}}})$, $\theta^{12,2}({{{f}}}_{
1},{{{f}}}_{
2}|{{{e_2}}})$, $\theta^{12,12}({{{f}}}_{
1},{{{f}}}_{
2}|{{{e_1,e_2}}})$, $\delta^{1,2}({{{e}}}_{
1}|{{{e'_2}}})$, $\delta^{1,2}({{{e}}}_{
1}|{{{e'_1,e'_2}}})$.

In the right graph of Figure \ref{figese3} there is contemporaneous independence between the two latent variables, i.e. $E_{1}(t) \independent  E_{2}(t)|E_{1}(t-1)$ that is equivalent to the zero restrictions on $\delta^{12}({{{e}}}_{
1},{{{e}}}_{
2})$, $\delta^{12,1}({{{e}}}_{
1},{{{e}}}_{
2}|{{{e'_1}}})$, $\delta^{12,2}({{{e}}}_{
1},{{{e}}}_{
2}|{{{e'_2}}})$, $\delta^{12,12}({{{e}}}_{
1},{{{e}}}_{
2}|{{{e'_1,e'_2}}})$. As for the left graph, the parameters  $\theta^{12}({{{f}}}_{
1},{{{f}}}_{
2})$, $\theta^{12,1}({{{f}}}_{
1},{{{f}}}_{
2}|{{{e_1}}})$, $\theta^{12,2}({{{f}}}_{
1},{{{f}}}_{
2}|{{{e_2}}})$, $\theta^{12,12}({{{f}}}_{
1},{{{f}}}_{
2}|{{{e_1,e_2}}})$ are null according to the local independence condition.

\begin{figure}[h]
\centering \resizebox{0.7 \textwidth}{!}
{\includegraphics{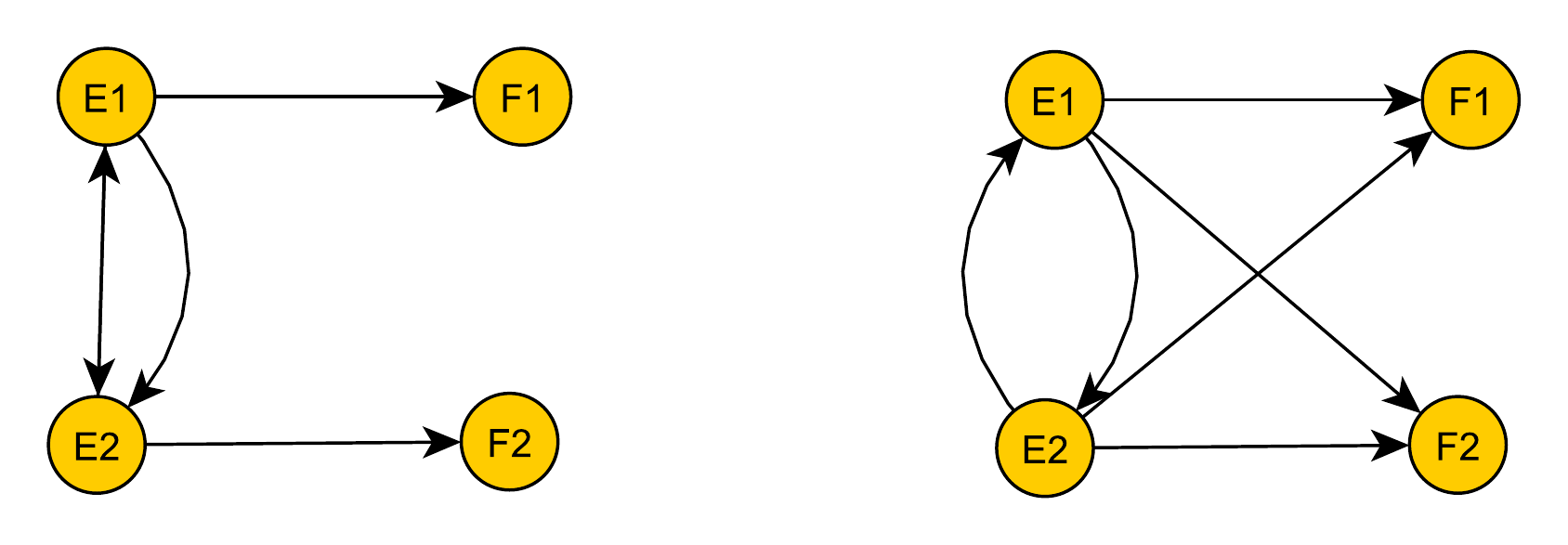}} \caption{Mixed-chain graphs whose Markov properties correspond to the constraints on marginal parameters  described in Example 4}
\label{figese3}
\end{figure}
\end{Example}

\subsection{Additivity hypotheses}
In the framework of hidden Markov models with several latent and
observable variables, other interesting hypotheses, that do not correspond to any conditional independence,  can be easily formulated. These hypotheses
reduce the number of parameters needed to parameterize the
transition and observation probabilities.
A useful restriction
that considerably simplifies  the observation model is the hypothesis of
\emph{additivity} of  the  effects of the latent variables on the marginal
interactions  of the observable variables. This marginal additive
dependence allows the interactions to
be expressed by the factorial expansion
\begin{equation}\label{teta2}
\eta^{\mathcal{P}}(\emph{\textbf{\text{f}}}_{
\mathcal{P}}|\emph{\textbf{\text{e}}}) =\theta^{\mathcal{P}}(\emph{\textbf{\text{f}}}_{
\mathcal{P}})+\sum_{k
\in \mathcal{U}}
\theta^{\mathcal{P},k}(\emph{\textbf{\text{f}}}_{
\mathcal{P}}|\emph{{\text{e}}}_{k}).
\end{equation}

Note that under this hypothesis, the parameters
 $\theta^{P,Q}(\emph{\textbf{\text{f}}}_\mathcal{P}|\emph{\textbf{\text{e}}}_\mathcal{Q})$ described in (\ref{teta}) are null if  $|Q|>1.$
 A similar additivity hypothesis can be used for the interactions of the latent model
\begin{equation}\label{teta3}
\lambda^{\mathcal{P}}(\emph{\textbf{\text{e}}}_{
\mathcal{P}}|\emph{\textbf{\text{e}}}') =\delta^{\mathcal{P}}(\emph{\textbf{\text{e}}}_{
\mathcal{P}})+\sum_{k
\in \mathcal{U}}
\delta^{\mathcal{P},k}(\emph{\textbf{\text{e}}}_{
\mathcal{P}}|\emph{{\text{e}}}'_{k}).
\end{equation}

 Another hypothesis is that of \emph{invariant association} corresponding to the constraints
  $\eta^{\mathcal{P}}(\emph{\textbf{\text{f}}}_{\mathcal{P}}|\emph{\textbf{\text{e}}})
=\theta^{\mathcal{P}}(\emph{\textbf{\text{f}}}_{
\mathcal{P}})$, if
$|P|>1$.  According to this hypothesis the
interactions between  observable variables do not depend on the states
of the latent variables.

Similarly, the constraints for the invariant association can be also imposed to the interactions of the latent model: $\lambda^{\mathcal{P}}(\emph{\textbf{\text{e}}}_{
\mathcal{P}}|\emph{\textbf{\text{e}}}')=\delta^{\mathcal{P}}(\emph{\textbf{\text{e}}}_{
\mathcal{P}})$, if
$|P|>1$.

\section{Examples}
In this section, we fit different MHMMs on two data sets. The EM algorithm used for estimating the models is described in Colombi and
Giordano, 2011, and implemented in the R-package hmmm by Colombi et
al., 2012.

  The data set of a soft-drink
company (Ching {\it et al.}, 2002, available also  in the R-package hmmm) consists of a one-year time series of daily sales  of
soft-drinks: lemon tea, orange juice and apple juice, all with categories: low, medium, high level. Changes in sale outcomes over
time can depend on time-varying unobservable factors and we consider an MHMM with two dichotomous latent variables to model these
data.

The marginal latent processes of the MHMM are denoted by $\mathbf{E}_{1},\mathbf{E}_{2}$, and the marginal observable components by  $\mathbf{F}_{{T}},$ $\mathbf{F}_{{O}}, \mathbf{F}_{{A}}$.

\begin{table}[h]\centering
\caption{Constrained latent and observation models for soft-drink
data}\label{cogiotab}
\renewcommand{\arraystretch}{0.9}
\setlength{\tabcolsep}{2.1 mm}
 \vspace{0.2 cm}
\medskip
\resizebox{12cm}{!} {
 \begin{tabular}{lccccccc}\hline
\emph{latent model} & \emph{obs. model}    &  $LRT$ & \emph{df} & \emph{p-value} & \emph{par} & \emph{loglike} & \emph{AIC}\\\hline
\emph{noGranger} &  \emph{saturated}   & 4.0705 & 4 & 0.3965 & 112 & -694.0009 & 1612.002
\\ \emph{noGranger} + \emph{ia}  & \emph{saturated }  &12.9629&7&0.0730 & 109 & -698.4471 & 1614.894\\
\emph{saturated} &  \emph{ci}   &33.0017&20&0.0337 & 96 & -708.4665 & 1608.933
\\ \emph{noGranger}  & \emph{ci}   &33.2346&24&0.0992 & 92 & -691.9656 & 1567.931\\
\emph{noGranger} + \emph{ia} &\emph{ci} &33.5379&27&0.1798 & 89 & -708.7346 & 1595.469\\
\emph{saturated} & \emph{ci} + \emph{ia} &76.2292&80&0.5987 & 36 & -730.0803 & 1532.161\\
\emph{noGranger} & \emph{ci} + \emph{ia} &77.4619&84&0.6795 & 32 &730.6966 & 1525.393\\
\emph{noGranger} + \emph{ia} &  \emph{ci} + \emph{ia} &77.9029&87&0.7467 & 29 & -730.9171 & 1519.834
\\ \emph{saturated} & \emph{ci} + \emph{ind}  & 89.6551&92&0.5497 & 24 & -736.7932 & 1521.586\\
\emph{noGranger} & \emph{ci} + \emph{ind}  &92.6434&96&0.5780 & 20 & -738.2874 & 1516.575\\
\emph{noGranger} + \emph{ia} & \emph{ci} + \emph{ind}  &94.1538&99&0.6189 & 17 & -739.0426 & 1512.085\\\hline
\end{tabular}}
\end{table}

Among others, we tested the hypotheses that the latent components $\mathbf{E}_{1}$ and $\mathbf{E}_{2}$ are marginally Markov chains and that the tea sales depend on $\mathbf{E}_{1}$ only, the orange and apple juices sales on $\mathbf{E}_{2}$ only.
That is we test for condition (\ref{c3}) of double Granger noncausality (in short \emph{noGranger}) for the latent variables:  $E_{1}(t)
\independent
{E}_{2}(t-1)|{E}_{1}(t-1)$ and $E_{2}(t)
\independent
{E}_{1}(t-1)|{E}_{2}(t-1)$ which ensure that $\mathbf{E}_{1}$ and $\mathbf{E}_{2}$ are univariate Markov chains and the restriction (\ref{c3b}) of conditional independencies for the observable variables given the latent chains: $ F_{T}(t)
\independent
{E}_{2}(t)|{E}_{1}(t)$ and $ F_{\{O,A\}}(t)\independent
{E}_{1}(t)|{E}_{
2}(t)$ that we will refer to as \emph{ci}.
These restrictions serve to assure that the marginal processes $(\mathbf{F}_{T},\mathbf{E}_{1})$ and $(\mathbf{F}_{O}\mathbf{F}_{A},\mathbf{E}_{2})$ are still hidden Markov models.

Additional hypotheses tested for the observable model are:  the  invariant association (\emph{ia}) mentioned in Section 4.2 meaning that the interactions of second and higher order of the observable variables do not depend on the latent states, i.e. $\eta^{\mathcal{P}}(\emph{\textbf{\text{f}}}_{
\mathcal{P}}|\emph{\textbf{\text{e}}})=\theta^{\mathcal{P}}(\emph{\textbf{\text{f}}}_{
\mathcal{P}})$ for $|\mathcal{P}|>1$, and the local independence (\ref{c2b}) (in short \emph{ind}) of all the observable variables given the latent chain  ${F}_{T}(t) \independent F_{O}(t) \independent F_{A}(t)
|{E}_{\{1,2\}}(t)$, while for the latent model we consider the hypothesis that there is invariant association (\emph{ia}) according to which the interactions of second order do not depend on the latent states at the past time occasions, i.e. $\lambda^{\mathcal{P}}(\emph{\textbf{\text{e}}}_{
\mathcal{P}}|\emph{\textbf{\text{e}}}')=\delta^{\mathcal{P}}(\emph{\textbf{\text{e}}}_{
\mathcal{P}})$, for $|\mathcal{P}|>1$.

Table~\ref{cogiotab} reports the likelihood ratio tests (LRT), degrees of freedom (df) and p-values for models restricted under the mentioned hypotheses against the unrestricted model, the number of parameters (par), the values of the log-likelihood function (loglike) and the Akaike criterion (AIC).\begin{figure}[h]
\centering \resizebox{0.3 \textwidth}{!}
{\includegraphics{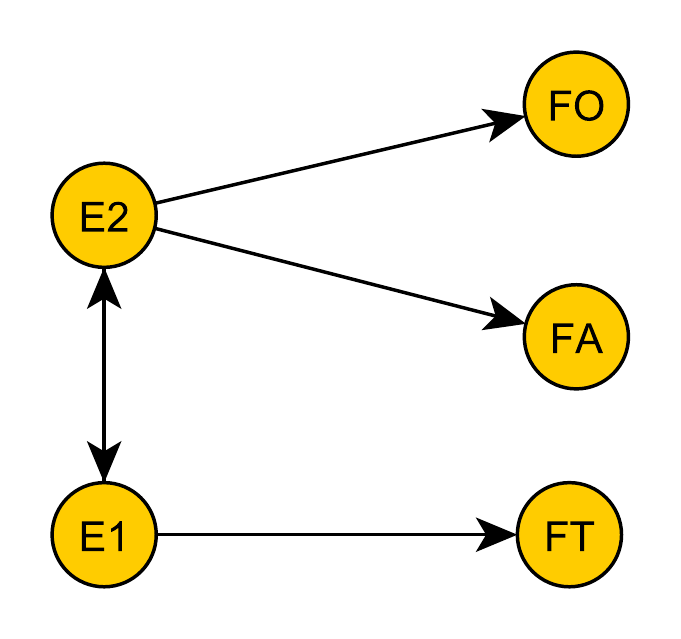}} \caption{Mixed-chain graph for soft-drinks data}
\label{bevande}
\end{figure}

The MHMM with the lowest AIC value is also the most parsimonious model (last row in Table~\ref{cogiotab}) which corresponds to the mixed-chain graph illustrated in Figure \ref{bevande}. It encodes the properties of double Granger noncausality and invariant association for the latent component of the MHMM, the local independence and the specific dependence of the observable variables on the latent chains for the observable component of the MHMM.

\vspace{0.5 cm}
The second data set, available at $http://archive.ics.uci.edu/ml/$, reports  daily measurements (from December 2006 to November 2010) of electric power consumption  in one household. The energy (in watt-hour of active energy) from the sub-meter 1 corresponds to the kitchen, containing mainly a dishwasher, an oven and a microwave; the energy from the sub-meter 2 is used for the laundry room, containing a washing-machine, a tumble-drier, a refrigerator and a light; energy from sub-meter 3 corresponds to an electric water-heater and an air-conditioner. Here, the energy consumptions registered over 4 years by the sub-meters 1, 2, and 3, are categorized in low and high levels according to whether the real measurements are under or over an established threshold, and ${\bf F}_1$, ${\bf F}_2$ and ${\bf F}_3$ indicate the resulting categorical time series.

The categorized consumptions for kitchen, laundry  and heater  are the manifest variables used as indicators of the true need of energy for eating, cleaning and heating/cooling which is  unobservable directly. The latent real request of energy, in fact, can be lower (higher) than the measured level due, for example, to waste (saving need).
Therefore, we model the energy data through an MHMM involving three latent variables with two states and three observable binary variables.

Several hypotheses has been considered to describe the relations among the observable series and the latent factors, but in the sequel we will focus only on those models which better perform in terms of interpretability, parsimony and fitting.

Let us start by considering the  saturated MHMM, i.e.\ without any restrictions, which attains $AIC = 5493.015$, $loglike = -2634.507$, $par = 112$.

Among the alternatives to the saturated model, a simple and intuitive linked  MHMM considers that each of the three observable series of energy consumptions  is an indicator of its specific latent variable and  the observable variables are locally independent.  Regarding  the latent model,  the three unobservable variables affect each others at the same time but each one depends only on the proper past (no Granger causality and no contemporaneous independence).  This linked  MHMM  ($AIC = 5621.251$, $loglike = -2778.626$, $par = 32$) cannot be preferred to the saturated model. Also the  coupled MHMM with every observable variable depending on its own latent variable and with contemporaneous independence and  Granger causality among the three latent variables ($AIC = 5615.272$, $loglike = -2777.636$, $par = 30$) is outperformed by the saturated model.

Unfortunately, although combined with other different hypotheses on the latent component of the MHMM, the model restricted under the  assumption of a specific effect of each latent variable on one and only one energy consumption series shows an unsatisfactory performance with an higher AIC value than that of the saturated model.

In the following  models, the effects of two or more latent variables on the same observable variable will be  considered additive as described in the expression (\ref{teta2}).

\begin{figure}[h]
\centering \resizebox{0.9 \textwidth}{!}
{\includegraphics{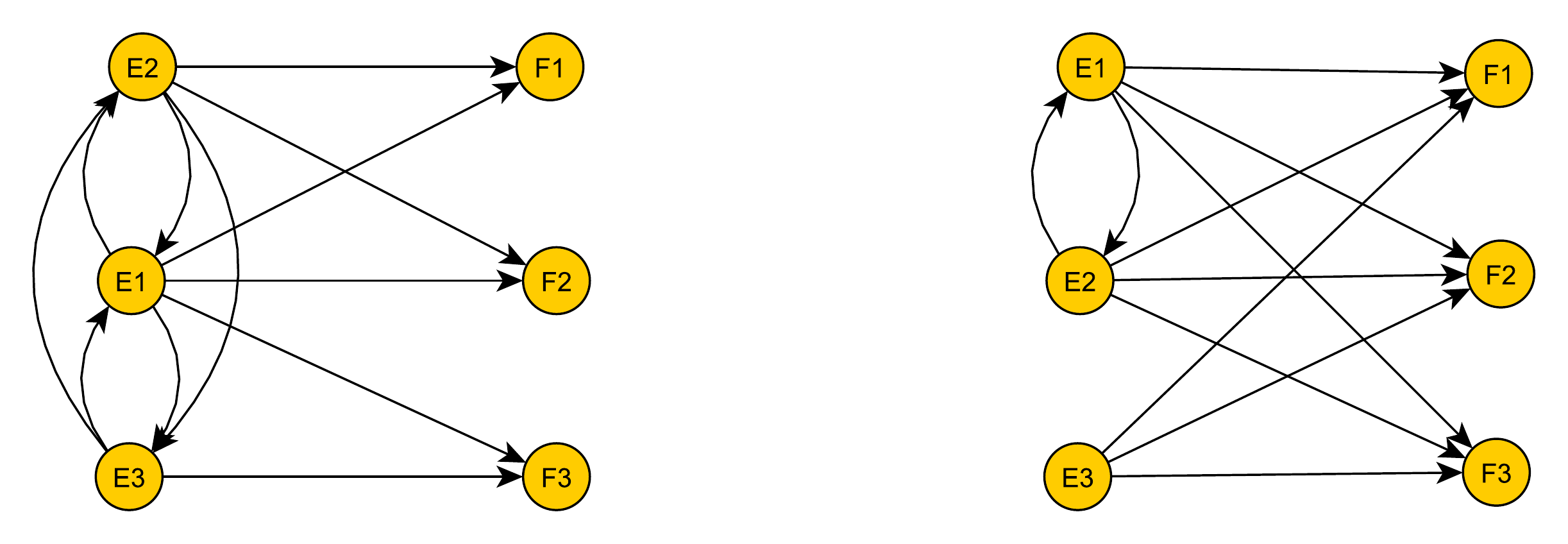}} \caption{Mixed-chain graphs for energy data}
\label{grafofinale}
\end{figure}

An interesting  model with a better fit assumes that: there is a specific effect of one latent variable on the energy consumptions  for kitchen and laundry, another latent variable influences the energy consumptions for the heater, and there is a generic effect since a third latent variable affects all the manifest variables. This  observation model is combined with the assumption that Granger causality and contemporaneous independence exist among all the latent variables. The model with these  restrictions has $AIC = 5470.029$, $loglike = -2702.014$, $par = 33$. Its corresponding mixed-chain graph is reported in Figure \ref{grafofinale} on the left.

The MHMM which seems the most suitable for representing the dynamics of the energy consumptions, among several models we fitted,  assumes that: the first two latent variables are G-caused reciprocally, the third one depends only on its past, and there is  contemporaneous independence among the three latent factors; moreover, there is local independence among the consumptions of the energy for eating, cleaning and air-conditioning which seem to depend on all the three latent variables ($AIC = 5459.77$, $loglike = -2707.883$, $par = 22$). Its  mixed-chain graph is reported in Figure \ref{grafofinale} on the right.
This model is nested in an MHMM which shows a similar fit  but admits that there is Granger causality among all the latent variables, the other hypotheses being equal ($AIC = 5467.374$, $loglike = -2697.687$, $par = 36$). When the two models are compared, the LRT ($LRT=20.4, df=14, p-value=0.118$) confirms that the most parsimonious model can be retained coherently with the AIC results.


\begin{thebibliography}{15}
\bibitem{BCF} Bartolucci, F., Colombi, R. and Forcina, A. (2007).
    An extended class of marginal
link functions for modelling contingency tables by equality and
inequality constraints.
     {\it Statistica Sinica}, {\bf 17},
     691\,--\,711.
\bibitem{BR} Bergsma, W. P. and Rudas, T. (2002).
     Marginal models for categorical data.
     {\it The Annals of Statistics}, {\bf 30},
     140\,--\,159.

  \bibitem{B}     Brand, M.,  Oliver, N. and   Pentland, A. (1997). Coupled hidden Markov models for complex action recognition. Proceedings of IEEE Computer Society Conference on Computer Vision and Pattern Recognition, 994\,--\,999.

\bibitem{capp} Capp\'{e}, O., Moulines, E. and Ryd\'{e}n T. (2005).
     {\it Inference in Hidden Markov
Models}.
     New York: Springer.

 \bibitem{Chi} Ching, W.K., Fung, E.S. and  Ng, M.K.(2002). A multivariate Markov chain for categorical data sequences and its
    applications in demand predictions. {\it IMA J. Manag. Math.}, \textbf{13}, 187\,--\,199.
\bibitem{COLGIO} Colombi, R. and Giordano, S. (2011).
     Testing lumpability for marginal discrete hidden Markov models.
     {\it Advances in Statistical Analysis}, {\bf 95}, 293\,--\,311.

\bibitem{COLGIOCAZ} Colombi, R. and Giordano, S. (2012).
Graphical models for multivariate Markov chains.
     {\it Journal of Multivariate Analysis}, {\bf 107}, 90\,--\,103.

     \item Colombi, R., Giordano, S. and Cazzaro, M. (2012).
     R-package {hmmm}: Hierarchical Multinomial Marginal Models.
     {\it http://CRAN.R-project.org/package=hmmm}.

\bibitem{Dr}   Drton, M. (2009).  Discrete chain graph models.  \emph{Bernoulli}, \textbf{15}(3),
 736\,--\,753.

\bibitem{Flo}  Florens, J.P., Mouchart,  M. and Rolin, J.M. (1993).   Noncausality and marginalization of Markov processes. \emph{Econometric Theory}, \textbf{9}, 241\,--\,262.




\bibitem{GlMc} Glonek,  G.J.N. and McCullagh, P. (1995).
    Multivariate logistic models. \emph{Jornal of the Royal Statistical Society, Ser. B Stat.
Methodol.}, \textbf{57},
533\,--\,546.

 \bibitem{K} Koski, T. (2001).  Hidden Markov Models for Bioinformatics. London: Kluwer Academic Publishers.

\bibitem{MACZUC} MacDonald, I.L. and Zucchini, W. (1997).
     {\it Hidden Markov and
  Other Models for Discrete-valued Time Series}.
    London: Chapman \& Hall.

\bibitem{MACZUC2} MacDonald, I.L. and Zucchini, W. (2009).
     {\it Hidden Markov Models for Time Series, An introdution using R}.
    London: Chapman \& Hall.

 \bibitem{ML}   Marchetti, G.M. and Lupparelli, M. (2011). Chain graph models of multivariate regression type for categorical data. \emph{Bernoulli}, \textbf{17}(3),  827\,--\,844.


    \bibitem{Ric}  Richardson, T. (2003). Markov properties for acyclic directed mixed
graphs. {\it Scandinavian Journal of Statistics}, \textbf{30},  145\,--\,157.

\bibitem{ZG}  Zucchini, W. and Guttorp, P. (1991). A hidden Markov model for space-time precipitation. \emph{Water Resourses Research}, \textbf{27}(8), 1917\,--\,1923.

\end{thebibliography}
\end{document}